# Methanol formation through reaction of low energy $CH_3^+$ ions with an amorphous solid water surface at low temperature

Y. Nakai [1], W. M. C. Sameera [2,3], K. Furuya [4], H. Hidaka [2], A. Ishibashi [2], and N. Watanabe [2]

[1] RIKEN Nishina Center for Accelerator-based Science, 2-1 Hirosawa, Wako, Saitama, 351-0198 Japan; nakaiy@riken.jp

[2] Institute of Low Temperature Science, Hokkaido University, N19W8, Kita-ku, Sapporo, Hokkaido 060-0819, Japan

[3] Department of Chemistry, University of Colombo, Colombo 00300, Sri Lanka

[4] National Astronomical Observatory of Japan, Osawa 2-21-1, Mitaka, Tokyo 181-8588, Japan

## Abstract

We have performed experimental investigations of methanol formation *via* the reactions of low-energy $CH_3^+$ ions with an amorphous solid water (ASW) surface at ~10 K. A newly developed experimental apparatus enabled irradiation of the ASW surface by several eV ions and detection of trace amounts of reaction products on the surface. It was found that methanol molecules were produced by low-energy $CH_3^+$ irradiation of the ASW surface and that hydroxy groups in produced methanol originated from water molecules in the ASW, as predicted in a previous theoretical study. Little temperature dependence of observed methanol intensity is apparent in the temperature range 12–60 K. *Ab-initio* molecular dynamics simulations under constant-temperature conditions of 10 K suggested that this reaction spontaneously produced a methanol molecule and an $H_3O^+$ ion, regardless of the contact point of $CH_3^+$ on the ASW surface. We have performed simulation with an astrochemical model under molecular-cloud conditions, where the reaction between $CH_3^+$ and $H_2O$ ice, leading to methanol formation, was included. We found that the impact of the reaction on methanol abundance was limited only at the edge of the molecular cloud (<1 mag) because of the low abundance of $CH_3^+$ in the gas phase, whereas the reaction between the abundant molecular ion ($HCO^+$) and $H_2O$ ice, which has not yet been confirmed experimentally, can considerably affect the abundance of a complex organic molecule. This work sheds light on a new type of reaction between molecular ions and ice surfaces that should be included in astrochemical models.

# 1. Introduction

Various kinds of neutral and ionic species have been found even in molecular clouds, despite their very low temperatures. In addition, recent improvements of capabilities of observing various targets have enabled the discovery of more species, including complex organic molecules (COMs) (e.g., Jørgensen et al. 2020). Many chemical processes have been proposed to explain how molecules evolve at each stage of an astronomical object. Among several types of reaction mechanisms in the gas phase, exothermic ion-molecule reactions have been considered to play an important role in chemical evolution, especially under low-temperature conditions, because no reaction barriers generally exist during formation of reaction intermediates, and thus an external energy input is not required. It has been revealed, however, that the abundances of some molecules such as hydrogen, water, and methanol molecules cannot be explained by only gas-phase synthesis. Accordingly, reactions of neutral radicals with neutral radicals or molecules on an icy grain surface have been proposed as the other essential reaction pathways, and their importance has indeed been confirmed for the above-mentioned molecules (e.g., Hama & Watanabe 2013; Watanabe & Kouchi 2002, 2008). In addition to the gas-phase and solid-phase reactions, processes directly triggered by interactions between gas-phase ions and ice mantles of grains could contribute to molecular synthesis. In this context, energetic-ion-bombardment experiments have been performed to simulate cosmic rays impinging on icy grains (e.g., Brown et al. 1982; Dartois et al. 2013; Boduch et al. 2015). The injection of energetic ions into ice mantles containing primordial molecules such as CO and $NH_3$ can produce a variety of molecules efficiently (Strazzulla 1998). However, little is known about the interactions between very-low-energy ions and dust-grain surfaces. In theoretical astrochemical models, molecular cations have been considered to simply recombine with negative charges (electrons) on grains (Walsh et al. 2013; Rimola et al. 2021). However, it is not clear how electrons exist on the realistic icy grains, e.g., degree of charge, localized or delocalized, and so on. Molecular cations may therefore not recombine immediately with electrons at the place where they land on the surface.

The reactions of very-low-energy ions with an ice surface have recently been theoretically proposed as new, non-energetic reaction pathways in addition to recombination (Woon 2011, 2015, 2021a, 2021b; McBride et al. 2014; Inostroza et al. 2019; Inostroza-Pino et al. 2021). In most of these proposed reactions, neutral species and $H_3O^+$ ions are spontaneously produced on the surface through reactions involving multiple water molecules. For example, quantum chemical calculations have predicted that the reaction of a $CH_3^+$ ion with a water ice surface could spontaneously produce a methanol molecule and $H_3O^+$ on the surface (Woon 2011). In contrast to the recent progress of theoretical studies of these processes, no relevant experimental study has been performed because of the technical difficulties associated with producing high-flux, low-energy ions and detecting trace amounts of products.

To investigate molecular syntheses by reactions of very-low-energy ions with ice, we have recently developed an experimental apparatus that enables irradiation of an ice surface with low-energy molecular ions and detection of trace of reaction products, and then, we selected the reaction of $CH_3^+$ with the ice surface as the target of the first experimental verification because no other reaction channels than methanol production were predicted. Using this apparatus, we experimentally confirmed for the first time that methanol molecules were produced through the reaction of low-energy $CH_3^+$ ions with a water–ice surface, as suggested theoretically. In addition, because secondary ions generated on/in the ice by ultraviolet (UV) photons and cosmic rays usually have kinetic energies in the range of 0.1–10 electron volts (eV) just after dissociative ionization of the parent molecule, such ions may induce non-thermal secondary reactions with molecules on/in ice in an energy range of a few electron volts. The present experiment is also worthwhile for considering secondary low-energy reactions following primary energetic processes.

We have also performed *ab-initio* molecular dynamics calculations for NVT (canonical) ensemble under constant temperature conditions of 10 K to illustrate that spontaneous methanol production can occur regardless of the initial configuration of $CH_3^+$-ice complex. This initial configuration dependence has not been explicitly shown in previous theoretical investigations although no other reaction channels have been suggested (Woon 2011, 2021b). Furthermore, to evaluate the impact of reactions between molecular ions and $H_2O$ ice on the chemistry in molecular clouds, we simulated the abundances of methanol and formic acid in molecular clouds by including their formation processes *via* the reactions of $CH_3^+$ and $HCO^+$ ions with $H_2O$ ice in the three-phase astrochemical model of Furuya et al. (2015).

## 2. Experiment

A schematic of our experimental equipment is shown in Figure 1. The equipment consisted of a main vacuum chamber, where a sample substrate was installed, a low-energy molecular ion source for $CH_3^+$ irradiation, and a $Cs^+$ ion source for detecting the product on ice. An aluminum substrate was mounted in a copper frame attached to the cold head of a He refrigerator at low temperatures (10–60 K). A large-bore rotation manipulator made it possible to rotate the substrate together with the He refrigerator so that the incident angle of the ions could be changed while cooling. A Pt-Co resistance temperature sensor with an accuracy of ±0.5 K was mounted on the copper frame of the Al substrate to monitor the temperature of the substrate. Two ceramic resistance heaters were fastened to the copper frame, which was used together with a temperature controller to maintain the temperature to within about ±1 K. An amorphous solid water (ASW) thin film was made on the substrate by introducing $H_2O$ (natural abundance water) or $D_2O$ (99.9%) vapor at a pressure of $\sim 1\times 10^{-6}$ Pa into the main vacuum chamber. The ASW generated by deposit of $H_2O$ ($D_2O$) vapor is hereafter referred to as $H_2O$-ASW ($D_2O$-ASW). Background base pressures of the main chamber were less than $\sim 4\times 10^{-8}$ Pa. Total exposure of water vapor was ~5 L (L; 1 langmuir = $1.33\times 10^{-4}$ pascal-seconds [Pa sec]).

The low-energy molecular ion source located at the front of the substrate was used for irradiation of the ASW surface with $CH_3^+$ ions. The $CH_3^+$ ions were generated by introducing $CH_4$ gas into an electron impact ionizer (B in Figure 1) and mass-analyzed through a Wien filter (C in Figure 1). The mass-selected $CH_3^+$ beam was decelerated with an electrostatic lens system (D in Figure 1) and impinged on the ASW surface for 1.5–3 hours at an incidence perpendicular to the surface plane. The intensities of $CH_3^+$, which were monitored on the Al substrate and its Cu frame, were ~0.45 nanoamperes (nA) and ~0.61 nA for the experiment using $H_2O$-ASW and $D_2O$-ASW, respectively. The energy distributions of $CH_3^+$ ions were measured by applying a retarding bias voltage to mesh plates that were inserted in the ion beam path just on the front of the substrate. The energy distributions of the ions were found to spread from ~3 eV to ~10 eV with a peak at around ~6.5 eV for $H_2O$-ASW experiments and from ~3 eV to ~11 eV with double peaks at ~6 eV and ~9 eV for $D_2O$-ASW experiments. The difference of $CH_3^+$ beam conditions between $H_2O$-ASW and $D_2O$-ASW experiments originated from slight difference of the setting of the molecular ion source. During $CH_3^+$ irradiation, the pressure increases in the main chamber due to leakage of $CH_4$ gas were typically $\sim 1\times 10^{-9}$ Pa or less.

The $Cs^+$ ion source (A in Figure 1) was equipped for detecting reaction products on the surface. This detection method, referred to as the reactive-ion scattering (RIS) method (Ishibashi et al. 2021; Kang 2011), has been developed mainly for detecting neutral and/or ionic species on metal and ice surfaces at relatively higher temperatures. Recently, the RIS method has been applied for the detection of absorbates on the ASW at around 10 K (Ishibashi et al. 2021). In the present study, we used a 40-eV $Cs^+$ ion beam with an intensity of 0.3–0.34 nA. The $Cs^+$ ions impinge at an incident angle 25° with respect to the substrate surface plane by rotating the Al substrate together with the He refrigerator. Some of the $Cs^+$ ions picked up neutral species, including water molecules, on the surface and were scattered as composite ions of $CsM^+$, where M was a neutral species picked up from the surface. Hereafter $CsM^+$ is referred to as a picking-up ion. The mass of M was obtained from a mass analysis with a quadrupole mass analyzer (QMA; E in Figure 1) of scattered $CsM^+$ ions, using 100% natural abundance of $^{133}Cs$. The axis of the QMA was set at 65° with respect to the substrate surface plane during the measurement of reaction products. Using the RIS method, we monitored the neutral species in the mass range 128–180 amu on the ASW surface before and after $CH_3^+$ irradiation. To confirm that no molecules were produced on the

ASW surface in the absence of $CH_3^+$, we also performed blank experiments under experimental conditions identical to the conditions during the $CH_3^+$ irradiation experiments, except that the $CH_3^+$ ions were blocked in the Wien filter and could not reach the surface of the ASW.

## 3. Results and Discussion

Figure 2(a) shows the mass spectra of picking-up ions in the mass range of 145–172 amu before and after 3-hour irradiation of the ASW surface by $CH_3^+$ ions at 11 K. The intensities of the mass-analyzed ions were normalized by the total yield of $Cs(H_2O)^+$ ions. Two strong peaks of $Cs(H_2O)^+$ (151 amu) and $Cs(H_2O)_2^+$ (169 amu) were always detected, regardless of ion irradiation. After $CH_3^+$ irradiation, a peak at 165 amu appeared. This peak could be assigned to $Cs(CH_3OH)^+$ and provided the first experimental evidence that $CH_3OH$ molecules were produced through $CH_3^+$ irradiation of the ASW surface. Because no peaks at 165 amu were apparent in the spectra before irradiation or in the blank experiment, it is ruled out that the detected $CH_3OH$ molecules after the irradiation originate from contamination from the low-energy molecular ion source and/or that those produced through the reactions induced by $Cs^+$ beam irradiation.

We also obtained small peaks at around 149, 150, 153, 168, and 171 amu after $CH_3^+$ irradiation. The peak at ~149 amu corresponded to the $Cs(CH_4)^+$. Methane molecules were due to leakage from the low-energy molecular ion source wherein $CH_4$ gas was used to produce $CH_3^+$ ions. Other peaks most likely corresponded to $Cs(OH)^+$ (150 amu), $Cs(H_2^{18}O)^+$ (153 amu), $Cs(OH)(H_2O)^+$ (168 amu), and $Cs(H_2O)(H_2^{18}O)^+$ (171 amu). The OH radicals might have been produced by protons that were contaminants in the $Cs^+$ ion beam or were produced at places other than the ASW surface and were adsorbed onto it. No products other than methanol were detected.

According to previous quantum chemical calculations, the simultaneous reactions of $CH_3^+$ with multiple water molecules in ice results in spontaneous formation of $CH_3OH$ and $H_3O^+$ (Woon 2011). The reaction can be described as follows:

$$CH_3^+ + nH_2O \text{ (ice)} \rightarrow CH_3OH + H_3O^+ \text{ (ice)} + (n-2)H_2O \text{ (ice)}. \qquad (1)$$

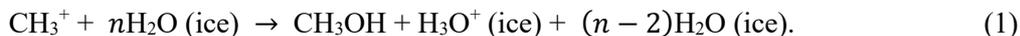

Production of methanol *via* this process consequently requires a hydroxy group from a water molecule on the ice surface. To confirm that hydroxy groups in methanol molecules detected here originated from water molecules in the ASW, we therefore performed experiments using $D_2O$-ASW at 12 K. In this case, $Cs(CH_3OD)^+$ with a mass of 166 amu should be produced. Figure 2(b) shows mass spectra in the mass range 147–175 amu of picking-up ions before and after 3-hour irradiation of the $D_2O$-ASW surface by $CH_3^+$ ions. A peak at 166 amu was apparent as well as peaks corresponding to water monomers and dimers picked up by $Cs^+$ ions, but there were no peaks at 165 amu corresponding to $Cs(CH_3OH)^+$. In addition, no peaks at 166 amu were apparent in the spectra before $CH_3^+$ irradiation and in the blank-experiment spectra. These results strongly support the hypothesis that the peak at 166 amu corresponded to $Cs(CH_3OD)^+$ and that hydroxy groups in methanol molecules originate from water molecules in the ASW. Note that the peaks of $Cs(H_2O)^+$ and $Cs(HDO)^+$ after $CH_3^+$ irradiation and in the blank-experiment spectra were more intense than those before $CH_3^+$ irradiation. The likely explanation is that $H_2O$ and $HDO$ molecules in the residual gas were adsorbed during $CH_3^+$ irradiation.

Figure 3(a) shows substrate temperature dependence of the methanol ($CH_3OD$) intensity integrated over the peak of methanol in the 3-hour $CH_3^+$ irradiation experiments with $D_2O$-ASW. Total amount of impinging $CH_3^+$ ions was estimated to match by ∼±5% for all temperatures shown in Figure 3(a) (12, 30, and 60 K). Little temperature dependence is apparent in the temperature range 12–60 K, which covers the temperature windows for the presence of $H_2$ and $CH_4$ molecules on the surface of the ASW. Methanol production during irradiation of $CH_3^+$ onto ASW is therefore not influenced by absorbates like $H_2$ and $CH_4$, which are thought to be dominant absorbances on the ASW surface around 10 K and 30 K. Thus, it is considered that the methanol production starting from the dissociation of adsorbed $CH_4$ unlikely occur.

Figure 3(b) shows the intensities of methanol as a function of the number of impinging $CH_3^+$ ions for the $H_2O$-ASW sample. The intensity did not increase linearly with the number of impinging $CH_3^+$ ions but instead gradually became saturated. This pattern might have resulted from local charging of the ASW (Tsekouras et al. 1998) within the area of $CH_3^+$ irradiation on the ice. We also observed a slight and gradual decrease of the beam current measured at the substrate during $CH_3^+$ irradiation. The beam current tended to recover when the bias voltage of the substrate was lowered. This behavior implies that the charging caused a slight distortion of the $CH_3^+$ orbits and that the $CH_3^+$ ions gradually avoided the charging region. Note that the influence of the charging depended on not only the density profile but also the distribution of the impinging energy of the $CH_3^+$ beam. These dependencies may have caused the detected yields to differ between the cases of $H_2O$-ASW and $D_2O$-ASW.

The actual kinetic energy of gas-phase $CH_3^+$ in interstellar matter is thought to be less than 0.05 eV (Woon 2021b). The range of kinetic energies in the present study was much higher. However, when the ions approach the ice surface, they are accelerated toward the surface by the attractive force due to the induced mirror charge in the ASW, even if no negative charges exist on the surface. The attractive potential, $qV_m$, that an ion with a charge state of $q$ feels is given by

$$qV_m = \frac{qe}{4\pi\varepsilon_0} \cdot \frac{1-\varepsilon_r}{1+\varepsilon_r} \cdot \frac{1}{2d} , \qquad (2)$$

where $e$ is the elementary charge, $\varepsilon_0$ is the vacuum electric permittivity, $\varepsilon_r$ is the relative permittivity of ASW, and $d$ is the distance between the ion and the ASW surface. Assuming $\varepsilon_r$ = 2.0 for ASW with a density of 0.63 g/cm$^3$ as discussed in Tsekouras et al. (1998), the values of the attractive potential voltage, $V_m$, are −1.2, −0.6, and −0.3 V at distances of 2, 4, and 8 Å (0 V at infinity), respectively, as shown in Figure 4. For an $\varepsilon_r$ = 3.0 for ice with a density of 0.93 g/cm$^3$, those values are −1.8, −0.9, and −0.4 V. For comparison, the mirror potential induced in a conductor (metal) is also shown. Furthermore, we also plotted the values of the relaxed potential energy for the reaction between a $CH_3^+$ ion and the ice surface with a quantum chemical calculation in Figure 4 and the corresponding values were estimated to be −4.2, −1.2, and −0.5 V at distances of 2, 4, and 8 Å, respectively (For details of the calculation, see Appendix A3.2). Consequently, $CH_3^+$ ions would have kinetic energies of around 1–4 eV when they collide on an ASW surface. Because the reaction dynamics are governed by the kinetic energy at the collision with the surface, the present experiments using ions with kinetic energies of several electron volts would approximately reproduce the actual reaction dynamics on the icy grain surface.

Here, we roughly estimate the ratio of detected methanol yields to total incident $CH_3^+$ ions. A simulation of $CH_3^+$ ion beam trajectories (SIMION$^{TM}$ version 8.0) indicated that about 70% of the $CH_3^+$ ions impinged within a rectangular area of 17 mm × 8 mm. Assuming that 70% of the $CH_3^+$ ions impinged uniformly within this rectangular area for 3 hours, we estimated the surface densities of the impinging $CH_3^+$ ions would be ~1.5 × 10$^{13}$ ions/cm$^2$ and ~2.1 × 10$^{13}$ ions/cm$^2$ for the cases of H$_2$O-ASW and D$_2$O-ASW, respectively. These surface densities would correspond to exposures of ~0.042 L (H$_2$O-ASW) and ~0.059 L (D$_2$O-ASW) for methanol gas at 300 K (see Appendix A2). Use of the relationship between exposure of methanol on H$_2$O-ASW and the intensity of methanol molecules picked up by Cs$^+$ ions (see Appendix A1) gives estimates of the ratios of detected methanol yields to total incident $CH_3^+$ ions of ~6% and ~11% for the cases of H$_2$O-ASW and D$_2$O-ASW, respectively. Therefore, assuming additionally that all impinging $CH_3^+$ are converted to methanol molecules as suggested by quantum chemical calculations (Woon 2011, 2021b), ~90% of methanol products might non-thermally desorb from the ASW surface at the present experiments. Considering that this estimation was based on rough assumptions and that kinetic energies of ions were higher than that in actual molecular clouds, the value of ~90% is probably near an upper bound. Nevertheless, we suppose that a large portion of produce methanol desorb non-thermally. The difference between the cases of H$_2$O-ASW and D$_2$O-ASW might have been due to the dependence of the influence of local charging on the distribution of the impinging energy of the $CH_3^+$ ions (*supra vide*).

**4. Theoretical investigation of dependence of methanol formation on initial configuration**

of $CH_3^+$-ice complex

*Ab-initio* molecular dynamics (MD) simulations, employing the atom-centered density matrix propagation (ADMP) method (Schlegel et al. 2002), were carried out for the 10 K constant temperature system (NVT ensemble) consisting of $CH_3^+$ and an ice cluster of 49 $H_2O$ molecules (Figure 5(a)), where the $CH_3^+$ was placed 3 Å above the ice cluster (Figure 5(b)). In a previous theoretical investigation, the influence of the initial configuration of $CH_3^+$-$(H_2O)_n$ complex was not explicitly shown (Woon 2011, 2021b). We therefore calculated trajectories starting from five initial positions of $CH_3^+$ on the surface of the ice cluster (Figure 5(c)). Details of the MD calculation are described in Appendix A3.1.

Figure 5(d) shows the temporal development of the bond distance between the C atom of the $CH_3^+$ ion and the O atom of the $H_2O$ close to the impact point. Figure A3 shows the snapshots of the trajectory 1 (see appendix A3.3). The equilibrium C-O bond distance of the $CH_3OH$ molecule is about 1.4 Å. In the case of trajectory **1**, the first contact of $CH_3^+$ with the ice surface thus occurs at around 100 femtoseconds (fs). After that time, the C-O bond vibration continues, while the amplitude of the oscillations gradually decreases with time, and $CH_3OH$ formation occurs by ~200 fs. The $CH_3OH$ is then stabilized on the ice, and the resulting $H_3O^+$ stays near the surface. Qualitatively similar results were obtained for trajectories **2**, **3**, **4**, and **5**: the first contact of $CH_3^+$ with the ice surface occurs at around 100 fs, and the C-O bond vibration continues around its equilibrium bond distance. Thus, the reaction between a $CH_3^+$ ion and ice spontaneously produced $CH_3OH$ and $H_3O^+$, regardless of the initial configuration of the $CH_3^+$.

## 5. Astrophysical implications

In astrochemical models, it is often assumed that a collision between a gas-phase molecular ion and a dust grain leads to the dissociative recombination of the ion rather than the formation of larger molecules (e.g., Walsh et al. 2013). As demonstrated in the present study, this is not always the case at least for the reaction between $CH_3^+$ and $H_2O$ ice. To study the impact of reactions between molecular ions and $H_2O$ ice on the chemistry in molecular clouds, we used the three-phase astrochemical model of Furuya et al. (2015), where a gas phase, an ice surface, and a bulk ice mantle are considered. The model includes gas-phase reactions, gas–grain interactions, and grain–surface reactions. The photodesorption yield of methanol per incident far-ultraviolet photon was set to $10^{-5}$ (Bertin et al. 2016) although this value is as large as its upper limit reported in Cruz Diaz et al. (2016), whereas 1% of the methanol formed by surface two-body reactions was assumed to be released into the gas phase by chemical desorption. For this work, we added

the reaction between $CH_3^+$ and $H_2O$ ice to our chemical model with the rate coefficient given by

$$k_{id} = \sigma_{id} v_{th} \theta(H_2O), \tag{3}$$

where $\sigma_{id}$ is the collisional cross section between a dust particle and an ion considering the Coulomb focusing (Spitzer 1941), $v_{th}$ is the thermal velocity of an ion, and $\theta(H_2O)$ is the surface coverage of the $H_2O$ ice. We assumed that 90% of the produced $CH_3OH$ was immediately released into the gas phase, whereas the remaining 10% remained on the surface, as implied by our experiments, because details of non-thermal desorption have not been well understood. Collisions between dust grains and molecular ions other than $CH_3^+$ were assumed to result in dissociative recombination. For physical conditions, we considered a one-dimensional model of a molecular cloud with a constant density of $10^4$ cm$^{-3}$ illuminated by the interstellar radiation field. The dust temperature was calculated via Equation 8 in Hocuk et al. (2017) with a lower bound of 10 K; e.g., 17 K at the edge of the cloud, but 10 K at a visual extinction ($A_V$) >~ 10 in magnitude (mag). The gas temperature was assumed to be the same as the dust temperature.

As shown in Figure 6(a), the reaction between $CH_3^+$ and $H_2O$ ice does not affect the $CH_3OH$ abundance significantly, except at the edge of the molecular cloud (<1 mag), where the gas-phase $CH_3OH$ abundance is very low (<$10^{-13}$) because of efficient UV photodissociation. The limited impact of the reaction in the UV-shielded regions (>1 mag) was due to the small abundance of $CH_3^+$ in our model (~$10^{-11}$), which was much lower than those of major molecular ions (e.g., $HCO^+$) and was due to the efficient formation of $CH_3OH$ by sequential hydrogenations of CO on grain surfaces (e.g., Watanabe & Kouchi 2002). Indeed, we found that addition of the reaction between $HCO^+$ and $H_2O$ ice, which leads to the formation of HCOOH, increased the HCOOH abundances in the gas and icy phases considerably (Figure 6(b)), where we assumed that 90 % of produced HCOOH is released into the gas phase, as the case of $CH_3^+$. Although that reaction has been suggested by previous quantum chemical studies (Woon 2011), it has not yet been confirmed by laboratory experiments. Further experimental studies of reactions between molecular ions and ASW surfaces are clearly needed to improve the understanding of the formation of large molecules at low temperatures.

## 6. Summary

We have experimentally investigated methanol formation through the reactions between low-energy $CH_3^+$ ions and an ASW surface at ~10 K. For the experiment, we have newly developed an experimental apparatus which enabled irradiation by several-eV $CH_3^+$ ions and detection of trace amounts of reaction products on the surface. Low-energy $CH_3^+$ irradiation of the ASW surface was found to produced methanol molecules on the surface, as a previous theoretical prediction (Woon 2011). The experiment using $D_2O$-ASW strongly suggested that hydroxy groups in produced methanol molecules originated from water molecules in the ASW. Little temperature dependence of detected intensities of methanol molecules is recognizable in the temperature range 12–60 K. It indicates that methanol production through $CH_3^+$ irradiation onto ASW is not influenced by absorbates like $H_2$ and $CH_4$. *Ab-initio* molecular dynamics simulations for NVT ensemble at 10 K suggested that the reaction of $CH_3^+$ with the ASW spontaneously produced a methanol molecule, regardless of the contact point of $CH_3^+$ on the ASW surface. Further, the reaction mechanism consists of (a) $CH_3^+$ ion interacting with the $H_2O$ molecule on ice, (b) C-O bond formation, giving rise to $CH_3OH$ and $H_3O^+$. The simulation with an astrochemical model has been performed under molecular-cloud conditions, where methanol production through the reactions between $CH_3^+$ and $H_2O$ ice was included. It was found that grate influence of the reaction on methanol abundance is limited only at the edge of the molecular cloud due to the low $CH_3^+$ abundance in the gas phase. In contrast, the reaction between the $HCO^+$, being more abundant in the gas phase, and $H_2O$ ice can markedly affect the abundance of formic acid. Further experimental investigations are necessary for reactions of low-energy molecular ions with ice surfaces.

**Acknowledgement**

This work was supported by JSPS KAKENHI Grant Numbers JP18H01273, JP17H06087, JP20H05847, JP21H05442, JP21H05416, JP22H00159. The authors grateful for partly support by the Grant for Joint Research Program of the Institute of Low Temperature Science, Hokkaido University (22K002). The authors thank the staff of Technical Division at Institute of Low Temperature Science at Hokkaido University for making the deceleration-lens system. Supercomputing resources at the Institute for Information Management and Communication at Kyoto University in Japan and Institute for Molecular Science (IMS) in Japan are acknowledged.

**Appendix**

*A1. Total normalized yield versus exposure of methanol*

We performed the detection of methanol artificially deposited on the ASW surface using the RIS method to obtain the relationship between the amount of detected methanol on the ASW surface and methanol exposure. First, an ASW film was made on the Al-substrate by introducing water vapor with an exposure of ~4 langmuirs (L) into the vacuum chamber. Then, methanol was deposited by exposing the ASW surface to methanol vapor, and then we used the RIS method to twice monitor neutral adsorbates with masses of 128–180 amu. We repeated methanol exposure on the same ASW film and RIS measurements two more consecutive times, i.e., methanol exposure and RIS measurements were repeated three times in succession. The methanol exposure for each deposit was in the range of 0.023–0.026 L. We thereby obtained the relationship between the total picked-up methanol yields normalized by picked-up water molecules in the first round of each of the RIS measurements as a function of total methanol exposure (Figure A1). This relationship was well described by

$$Y = 0.329 \times \left[1 - \exp\left(-\frac{D}{0.193}\right)\right], \quad (A1)$$

where $Y$ is the total normalized yield of picked-up methanol, and $D$ is the total methanol exposure in L. Influences of sputtering by $Cs^+$ ions were estimated based on the decrease of the picked-up methanol intensity between the first and second round of each RIS measurement. Total methanol exposures on the horizontal axis of Figure A1 were estimated by taking account of this decrease by sputtering.

### A2. Conversion between exposure and surface density of methanol

We converted exposure of methanol to surface density as follows: the flow $F$ (molecules/cm$^2$/sec) of gaseous methanol colliding onto the ASW surface at pressure $p$ (Pa) is given by

$$F = \frac{p}{\sqrt{2\pi m k_B T}}, \quad (A2)$$

where $m$ is the mass of methanol in kg, $k_B$ is the Boltzmann constant (1.38×10$^{-23}$ J/K), and $T$ (K) is the temperature of gaseous methanol. Assuming that the sticking coefficient is unity at substrate temperatures below 60 K and that $T$ = 300 K, the exposure of 1 langmuir (L) (1 L = 1.33×10$^{-4}$ Pa × 1 sec) corresponds to 3.58×10$^{14}$ molecules/cm$^2$.

### A3. Computational methods

### A3.1. Ab-initio molecular dynamics simulations

A periodic ice slab structure (25 × 25 × 25 Å) containing 493 $H_2O$ molecules (density 0.94 g/cm$^3$) was prepared and optimized using the AMOEBA09 (Ponder & Case 2003; Ren & Ponder 2002, 2003) force field in the TINKER program (version 8.10) (Kundrot et al. 1991; Ponder & Richards 1987). Molecular dynamics (MD) simulations, a 5 ps NVT ensemble at 300 K followed by a 5 ps NPT ensemble, were then performed at 300 K. The Andersen thermostat was used for the NVT simulation, whereas the Berendsen barostat was applied for the NPT simulation. The time step of 1.0 fs was used for both the NVT and NPT simulations. Then, the system temperature was set to 10 K, and 5 ps NVT followed by 5 ps NPT simulations were performed. The resulting structure was fully optimized using the AMOEBA09 force field, and a cluster model of 49 $H_2O$ molecules was prepared (see Figure 5(a)).

The ice cluster model was fully optimized using the ωB97X-D (Chai & Head-Gordon 2008) functional and 6-31G(d) (Ditchfield et al. 1971; Hariharan & Pople 1973; Hehre et al. 1972) full electron basis sets for all atoms, as implemented in the Gaussian16 program (version C.01) (Frisch et al. 2016). Cartesian coordinates of the optimized structure are given in Table 1. Then, a $CH_3^+$ was placed 3 Å above the ice cluster (see Figure 5(b)) and optimized the system keeping all $H_2O$ molecules frozen. After that, *ab-initio* molecular dynamic (MD) simulations for the NVT ensemble of $CH_3^+$ and the ice cluster, employing the ADMP (Schlegel et al. 2002) method, were performed. We have chosen the ADMP for *ab-initio* MD, as this method give equal functionality to Born-Oppenheimer molecular dynamic (BOMD) simulations at a relatively low computational cost (Schlegel et al. 2002). The ωB97X-D functional and 6-31G(d) basis sets were used for the ADMP simulations with a velocity-scaling thermostat. The time step was 0.1 fs, the fictitious mass was 0.1 amu, and the maximum simulation time was 400 fs except trajectory **1** (1000 fs). The total initial nuclear kinetic energy of the system was set to 0.1952 eV (0.00129 eV/atom) so that the value per atom corresponds to $3k_BT/2$ at the initial temperature of $T = 10$ K, and the target temperature for nuclear motion of the whole system was kept at 10 K with an allowed deviation from the target temperature of ±3 K.

*A3.2. Calculation of the relaxed potential energy surface for the reaction between a $CH_3^+$ ion and the ice surface*

To study the relaxed potential energy surface for the reaction between $CH_3^+$ and ice surface, an ice cluster model consisting of 240 $H_2O$ molecules was prepared (density 0.94 g/cm$^3$). The ONIOM(ωB97X-D/6-31G(d):AMOEBA09) method, as implemented in the PyQM/MM interface (Rathnayake et al. 2022), was used for structure optimizations, where 50 $H_2O$ molecules are in the QM region and the remining 190 $H_2O$ molecules are in the MM region. A $CH_3^+$ ion was then placed at a distance of 10.0 Å above the surface (Figure A2), and a relaxed potential energy surface was calculated. The final potential energy of the optimized structures was calculated as the single-point energy, employing the ωB97X-D/6-31G(d) method for all atoms in the ice cluster model.

Values of the obtained potential energy at distances of 5–10 Å were fitted by Equation 2 to estimate the value at infinite distance because the potential energy is strongly influenced by the details of locations of atoms in the ice as the ion approaches the ice surface. The results of the fitting showed that the relative permittivity of the model ice was probably $\varepsilon_r = 3.2 \pm 0.6$, which is consistent with a previously reported value of $\varepsilon_r = 3.0$ for ASW with a density of 0.93 g/cm$^3$ (Tsekouras et al. 1998). In Figure 4, we show the differences from the value at infinite distance as the potential energies so as to set the value at infinite distance to zero. The uncertainties of these difference values, i.e., potential energies, originating from the fitting procedure, are estimated to be about ± 0.1 eV.

*A3.3 Snapshots of the trajectory **1***

Figure A3 shows the snapshots of the trajectory **1** at every 100 fs from 0 to 1000 fs. At around 100 fs, $CH_3^+$ contact with a $H_2O$ molecule on the ice surface. Then, the C-O bond length decreases and reaches the equilibrium bond distance around 100-200 fs, and $CH_3OH$ formation is completed around 200 fs. Then, $CH_3OH$ is stabilized, and the resultant $H_3O^+$ ion is still nearby. Other trajectories showed a qualitatively similar picture. In actual experimental condition, some of $H_3O^+$ ions would move towards the Al substrate through proton transfer between $H_3O^+$ and $H_2O$ in the ice.

**Figures captions**

**Figure 1**. Schematic of the experimental setup. A: $Cs^+$ ion source; B: Electron-impact ionizer of the low-energy molecular ion source; C: Wien filter used for mass selection; D: Deceleration lens system; E: Quadrupole mass analyzer (QMA); F: Aluminum substrate on which the ASW film is formed. The vacuum chambers of the electron-impact ionizer, of the Wien filter, and of the main chamber were differentially evacuated through beam slits with 2-mm-diameter apertures.

**Figure 2**. (a) Mass spectra of picking-up ions in the mass range 145–173 amu after a 3-hour irradiation of the $H_2O$-ASW surface by $CH_3^+$ ions (black solid line), before the irradiation (red dashed line), and for the blank experiment (blue dashed line). (b) Mass spectra of picking-up ions from the $D_2O$-ASW surface by the RIS method in the mass range 147–175 amu after a 3-hour irradiation (black solid line), before the irradiation (red dashed line), and for the blank experiment (blue dashed line).

**Figure 3**. (a) Substrate temperature dependence of the intensity of methanol ($CH_3OD$) in the $CH_3^+$ irradiation experiments using $D_2O$-ASW (3-hour irradiation of $CH_3^+$ ions). The intensity is normalized by the total yield of $Cs(H_2O)^+$ ions. Red arrows show the temperature ranges for thermal desorption of $\lesssim 1$ monolayer-thick $H_2$ and $CH_4$ (Amiaud et al. 2015; Smith et al. 2016) (b) Total normalized intensities of methanol for the RIS measurement as a function of the total number of impinging $CH_3^+$ ions. The red solid circle is for 3-hour irradiation of $CH_3^+$ ions on the $H_2O$-ASW surface. Open and solid blue circles are for 1.5-hour irradiation and subsequent further 1.5-hour irradiation (total 3-hour irradiation) on the $H_2O$-ASW surface, respectively. The purple dashed line shows the fitted curve given by the equation $0.0056 \times [1 - \exp(-N_{inc}/(2.0 \times 10^{13}))]$, where $N_{inc}$ is the total number of impinging $CH_3^+$ ions.

**Figure 4**. Attractive potential as a function of the distance from the surface when ions approach the ASW surface feel. The black solid line shows the potential when the relative permittivity of the ASW is 2. The red solid line shows the relationship when the relative permittivity of the ASW is 3. The blue dotted line shows the relationship for a conductor for comparison. The violet solid circles are the values obtained via quantum chemical calculations.

**Figure 5**. (a) Ice cluster model employed in this study. (b) Side view of a $CH_3^+$ ion on the ice cluster. (c) Top view of a $CH_3^+$ ion on the ice cluster and its initial positions (1–5, yellow color circles) for the ADMP trajectory calculations. (d) Temporal development of the bond distance between the C atom of the $CH_3^+$ ion and the O atom at the impact point of the five ADMP

trajectories for the reaction between the $CH_3^+$ and $H_2O$ that leads to the formation of $CH_3OH$. The numbers from 1 to 5 in the legend correspond to the five initial positions of the $CH_3^+$ in (c), respectively.

**Figure 6**. $CH_3OH$ (a) and $HCOOH$ (b) abundances as functions of the visual extinction ($A_V$) predicted by the astrochemical model of a molecular cloud. Red lines indicate the model with the reaction between $H_2O$ ice and $CH_3^+$ (or $HCO^+$), leading to the formation of $CH_3OH$ (or $HCOOH$), whereas black lines indicate the model where collisions between molecular ions and dust grains lead to dissociative recombination. (c) The abundances of relevant species, $H_2O$ in ice, $CH_3^+$, and $HCO^+$.

**Figure A1**. Total picked-up methanol intensity normalized by picked-up water molecules in the first round of each RIS measurement as a function of total methanol exposure. Red dashed line shows the fitted curve given by Equation A1.

**Figure A2**. Ice cluster model with the $CH_3^+$ ion on it, where "*r*" is the reaction coordinate. The ONIOM-high layer is shown in the form of "ball and stick", and the ONIOM-low layer is shown in the form of "wireframe".

**Figure A3.** The snapshots of the trajectory **1** at every 100 fs from 0 to 1000 fs.

Figure 1

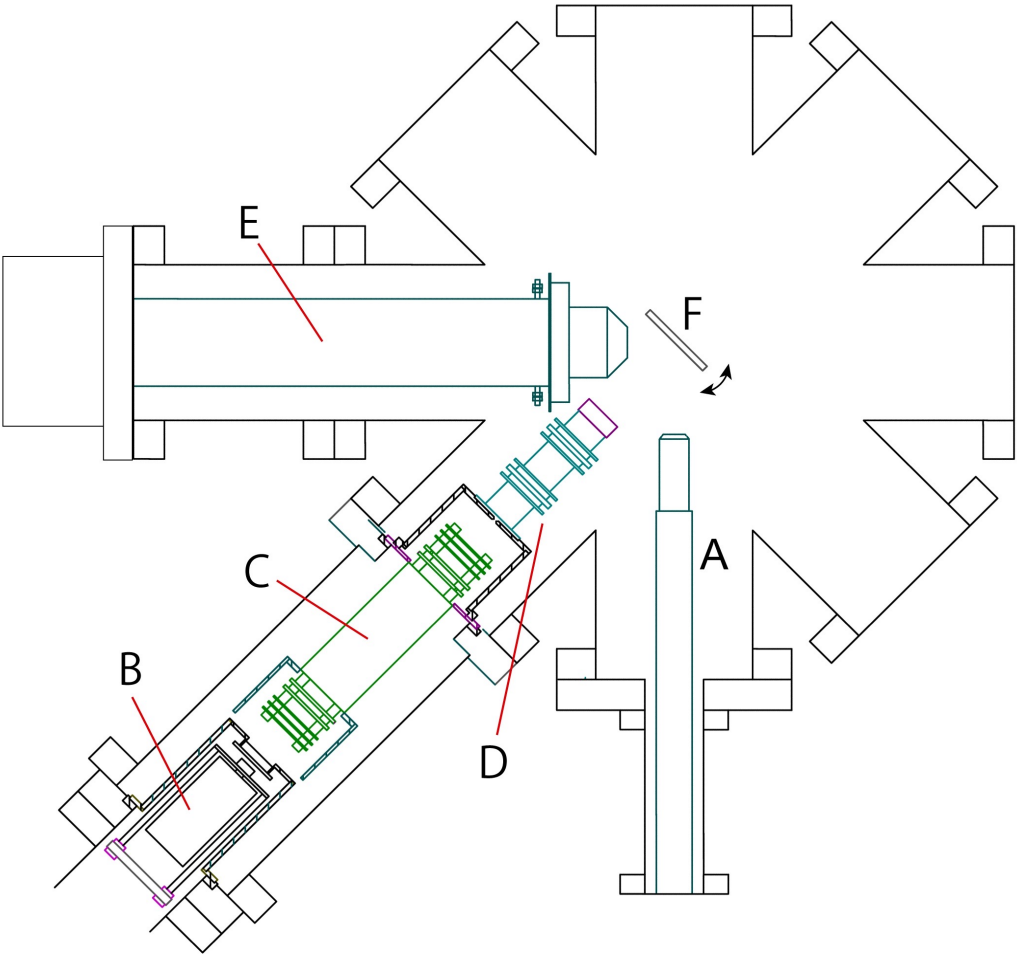

Figure 2

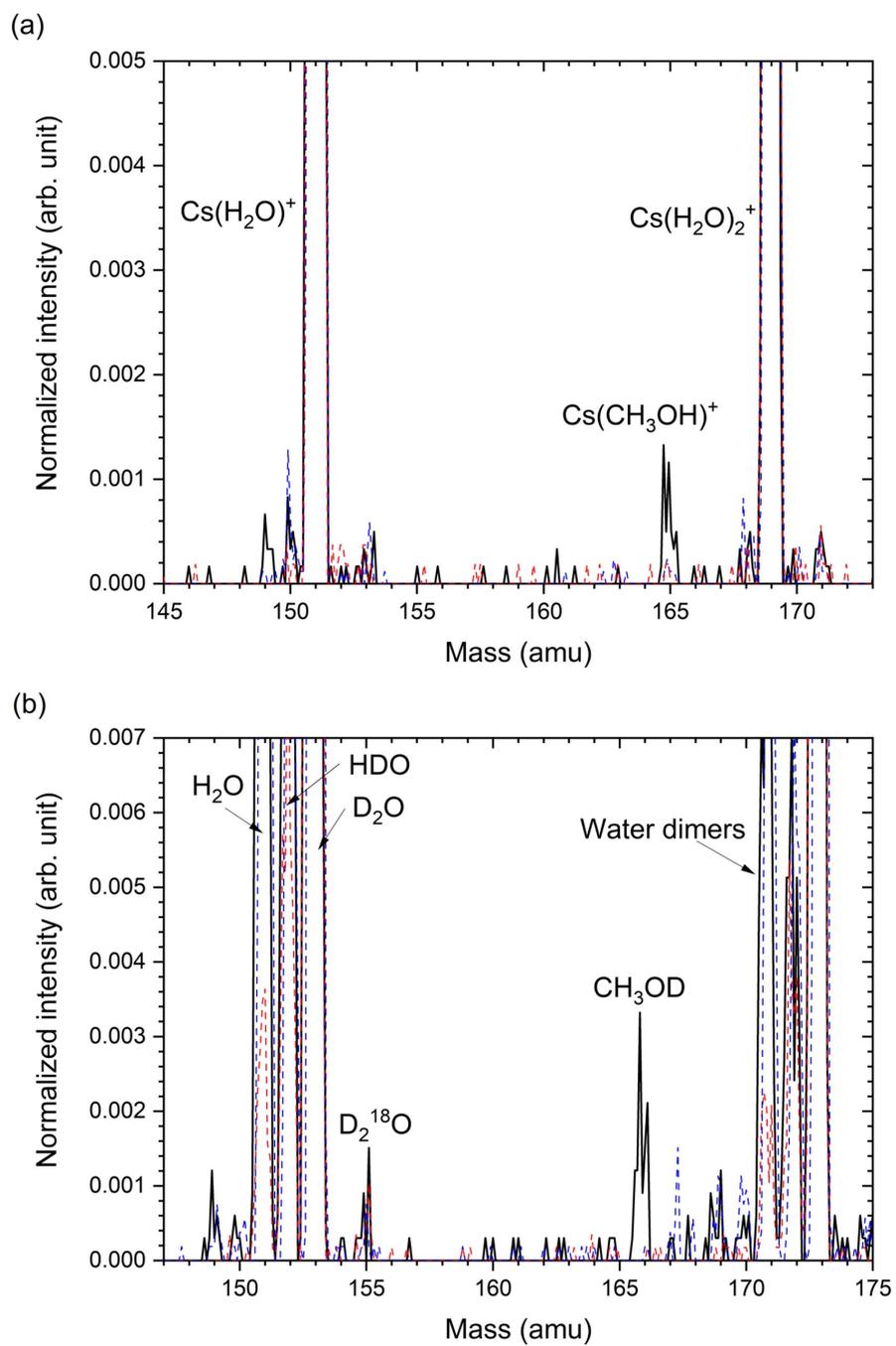

Figure 3

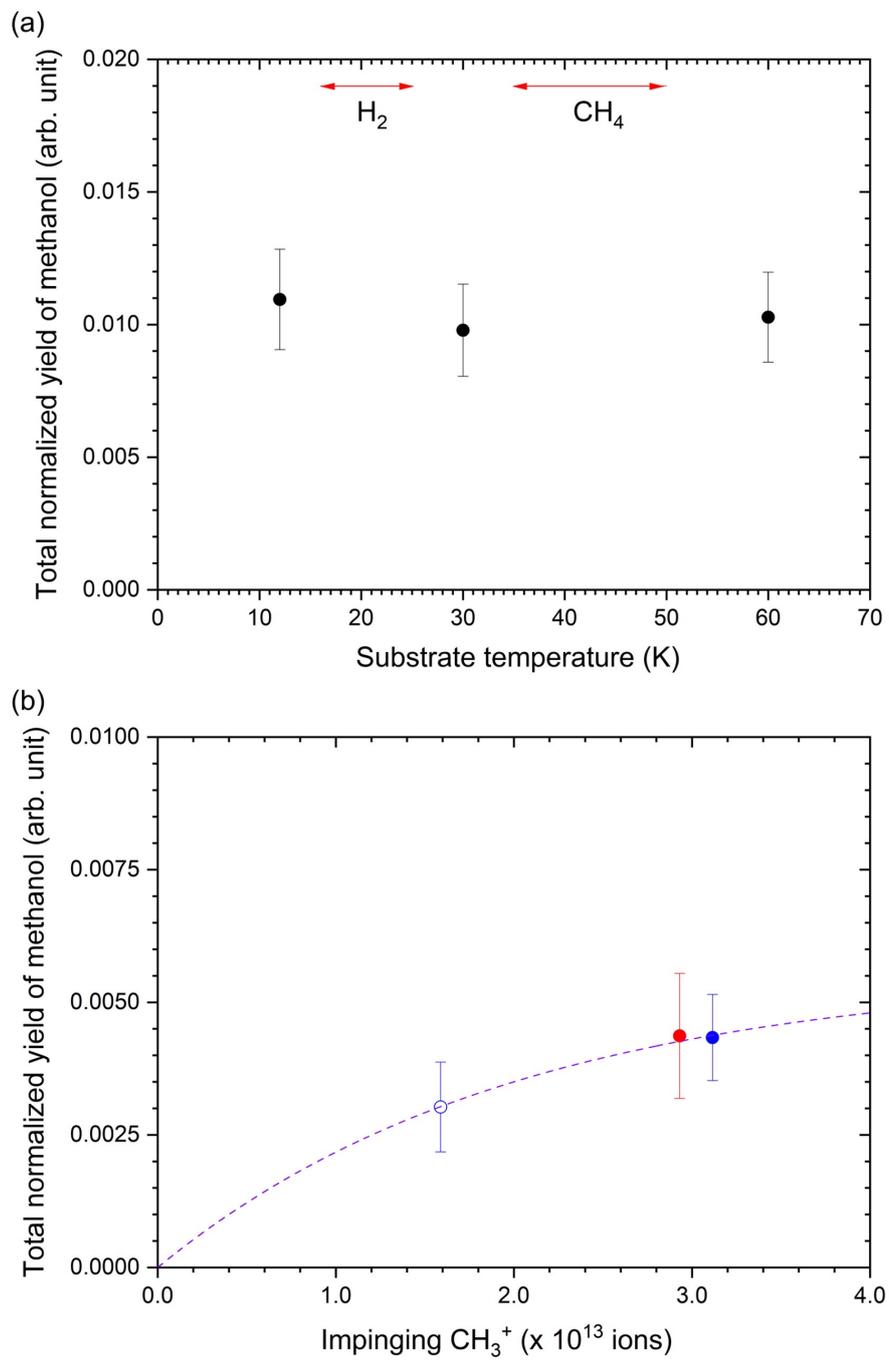

Figure 4

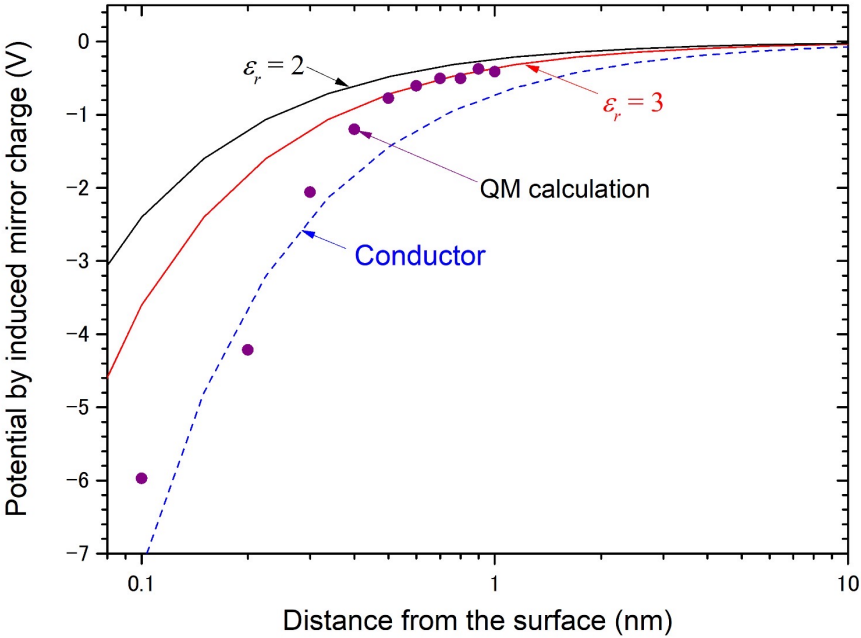

Figure 5

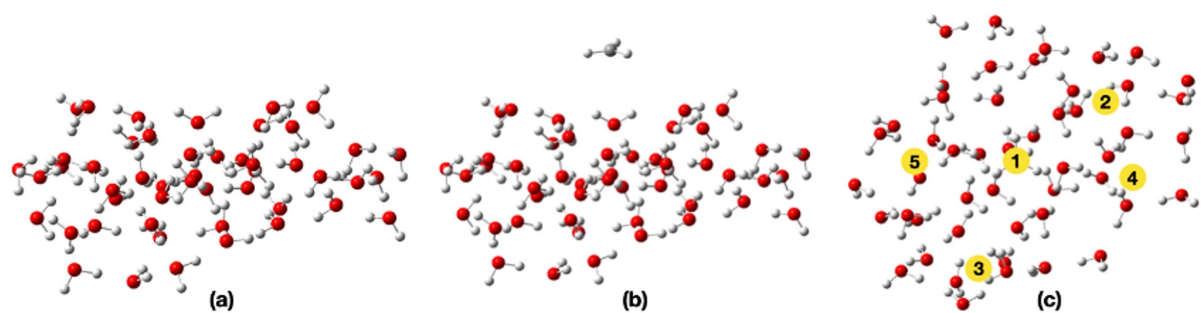

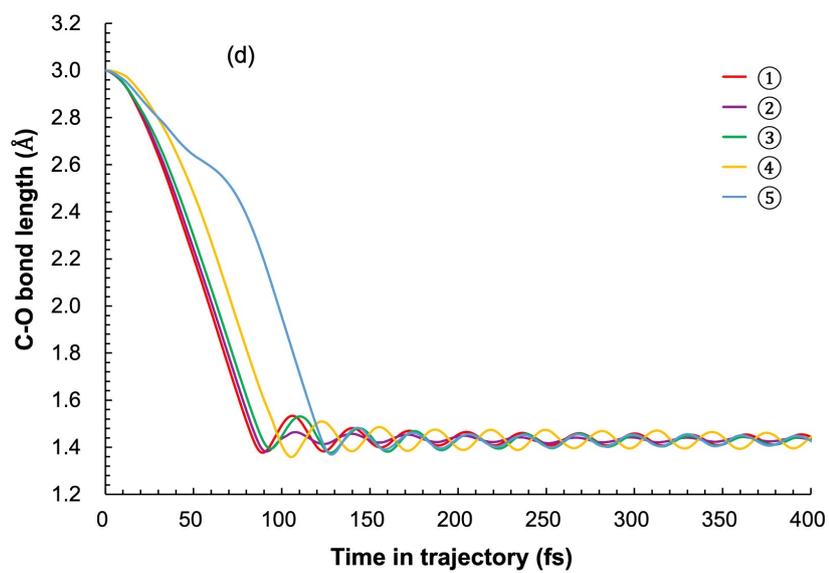

Figure 6

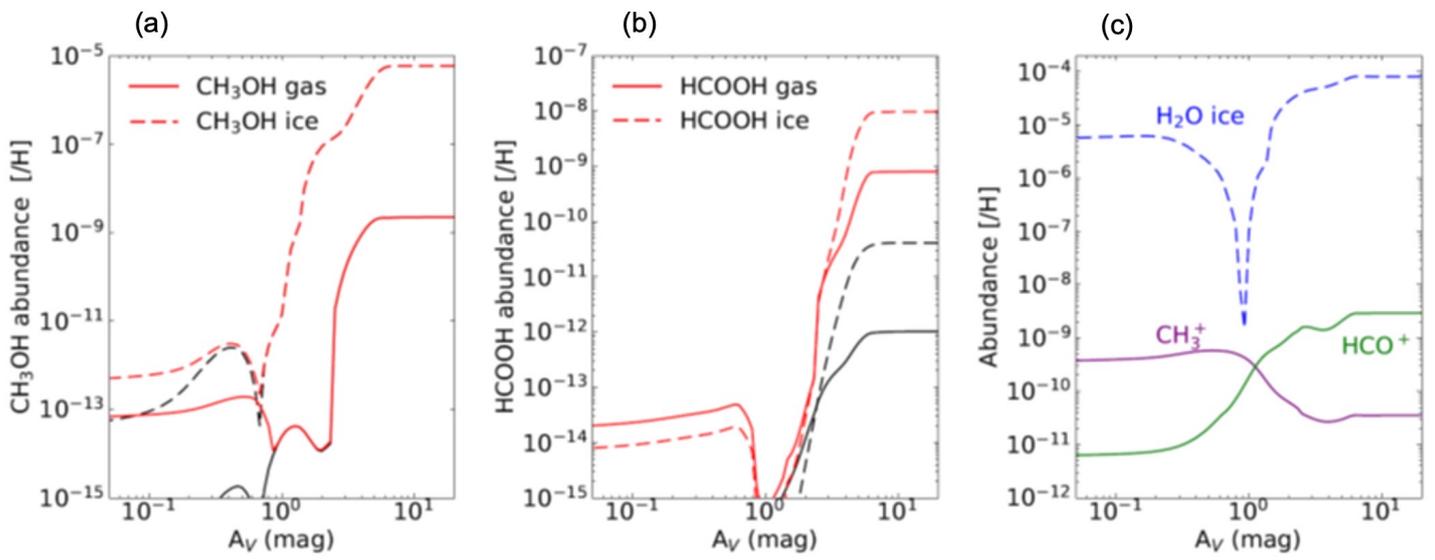

Figure A1

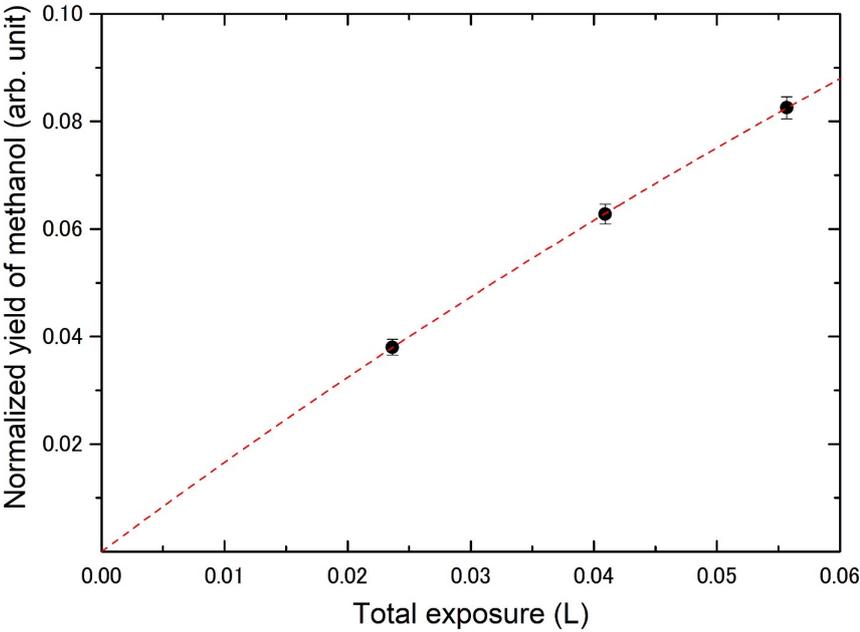

Figure A2

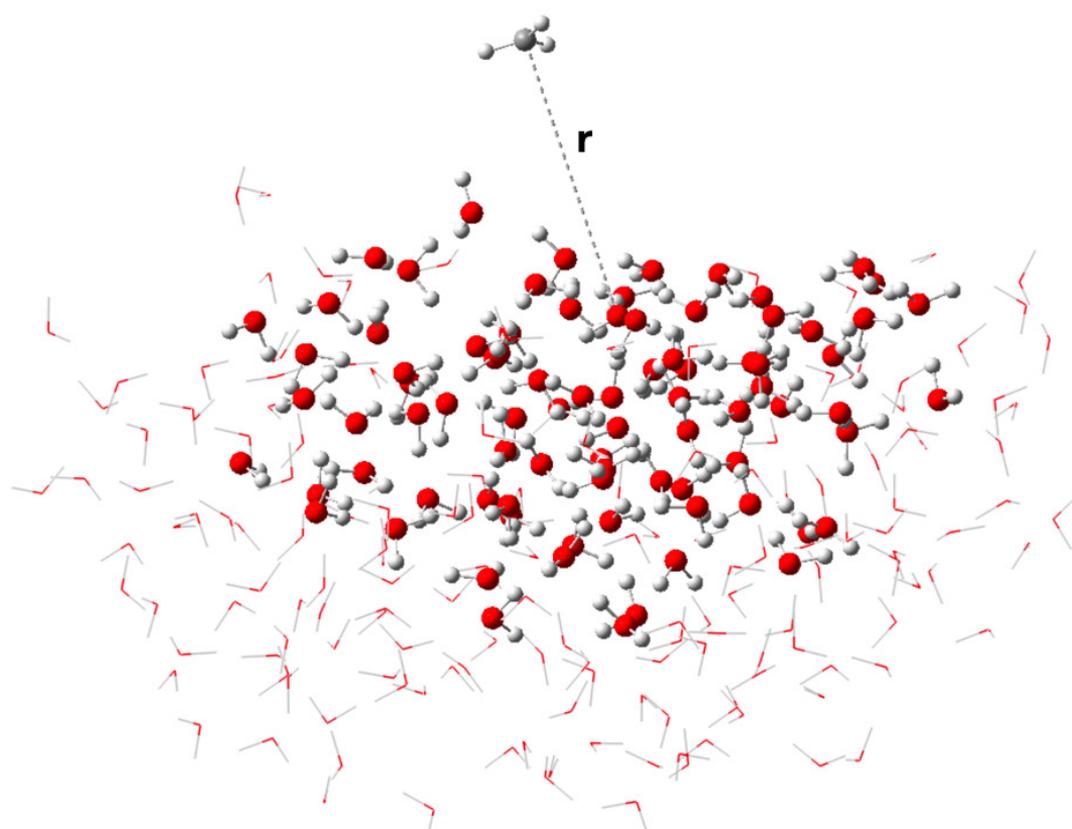

Figure A3

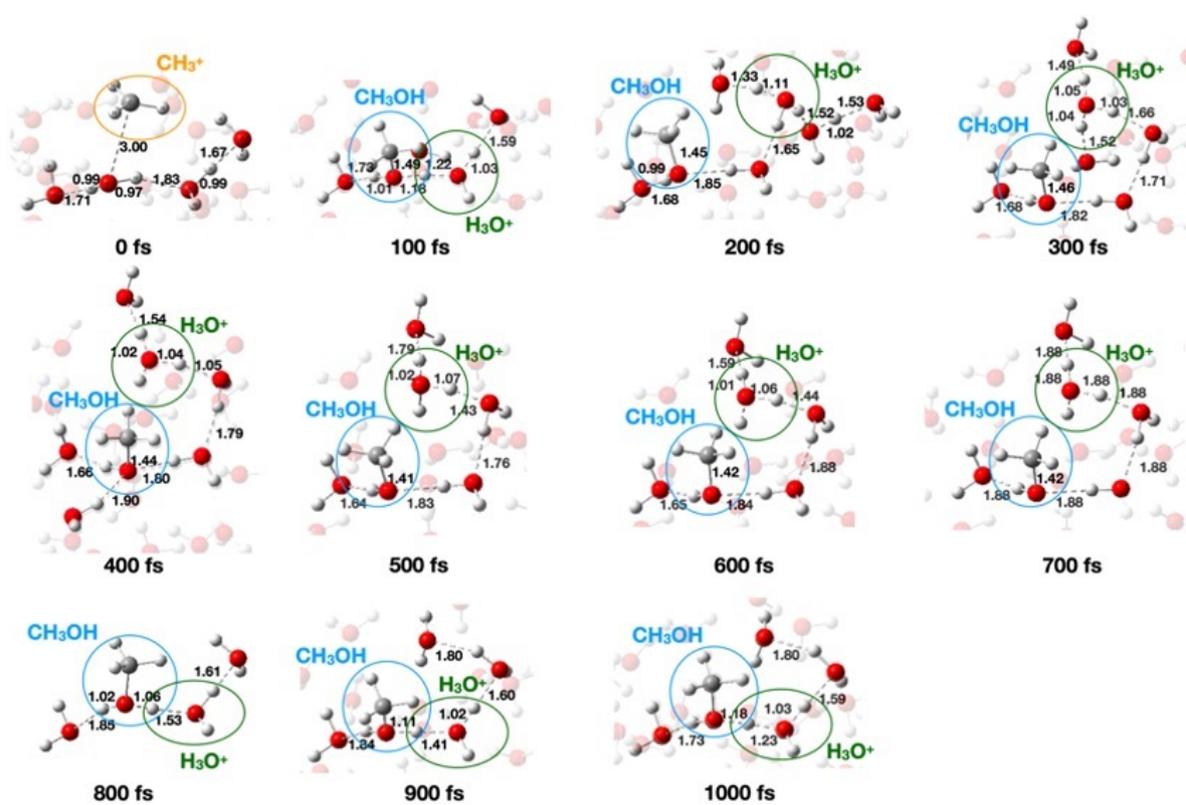

| 1 | **Table 1: Cartesian coordinates of the ice cluster used in this study (Å).** |
|---|---|

| | | | | |
|---|---|---|---|---|
| 3  | H | 5.25298541  | -2.76537520 | 4.60618526  |
| 4  | H | 3.92164031  | -2.27843317 | 3.94735122  |
| 5  | H | 1.69263915  | -2.26922817 | 4.08792123  |
| 6  | H | 2.26478419  | -0.84675907 | 3.92338321  |
| 7  | H | -0.58271001 | -2.57154619 | 3.30859417  |
| 8  | H | -0.19392698 | -3.63759027 | 4.37372325  |
| 9  | H | -8.29544857 | 2.90006220  | 2.76747913  |
| 10 | H | -6.96141747 | 2.90880520  | 1.95146307  |
| 11 | H | 0.51699207  | -4.35340332 | 2.23435409  |
| 12 | H | 1.38185613  | -4.35438632 | 0.94195600  |
| 13 | H | 5.87894245  | -2.41553018 | 0.79271799  |
| 14 | H | 6.00526246  | -2.24734617 | 2.32474310  |
| 15 | H | 6.35632949  | -5.71561042 | -0.84821313 |
| 16 | H | 4.80151838  | -5.51013940 | -0.90558013 |
| 17 | H | 4.68505137  | -4.04098530 | 2.68529013  |
| 18 | H | 5.22301541  | -4.65664334 | 1.33625603  |
| 19 | H | 3.47342928  | -6.18486145 | 1.73076206  |
| 20 | H | 1.96328317  | -6.34556946 | 1.58292105  |
| 21 | H | -8.29367557 | -2.37935418 | -0.81446513 |
| 22 | H | -6.80505246 | -2.20568316 | -0.44337410 |
| 23 | H | -2.18528713 | 3.91485628  | 2.01152208  |
| 24 | H | -2.15427812 | 2.50527917  | 2.67141412  |
| 25 | H | -2.65821816 | -1.42441511 | 2.44271311  |
| 26 | H | -1.34957807 | -0.57245905 | 2.42280411  |
| 27 | H | 1.51811414  | -0.39044303 | 2.03818008  |
| 28 | H | 2.58171622  | 0.69475204  | 2.22620309  |
| 29 | H | 0.13238604  | 1.08042107  | 2.46859011  |
| 30 | H | -0.86179303 | 1.42651210  | 1.31713603  |
| 31 | H | -8.14006655 | 2.48306217  | -0.96299314 |
| 32 | H | -8.31641857 | 3.41596824  | 0.20708095  |
| 33 | H | 2.27441819  | 6.45744346  | 0.71311598  |
| 34 | H | 3.78038830  | 6.82468449  | 0.76462199  |
| 35 | H | -5.55239237 | 4.51336732  | 0.09673494  |
| 36 | H | -6.69062745 | 4.38169031  | -0.92295713 |

| 37 | H | -7.94810754 | -0.26613402 | -0.03863307 |
| 38 | H | -8.08665455 | 1.06936207 | 0.82977299 |
| 39 | H | 3.51241428 | 3.60073525 | -0.53106211 |
| 40 | H | 3.39467228 | 4.56017032 | 0.73798599 |
| 41 | H | -0.58025801 | -5.39191239 | 0.28201895 |
| 42 | H | -0.52036101 | -5.28895539 | -1.27425316 |
| 43 | H | -0.36049100 | 4.01737128 | 0.88728300 |
| 44 | H | -1.40007307 | 4.76034034 | 0.03813494 |
| 45 | H | 7.60914658 | -1.37176810 | 0.47196497 |
| 46 | H | 6.98289253 | -1.35573810 | -0.89644313 |
| 47 | H | 6.97262253 | 0.65396804 | -0.22916508 |
| 48 | H | 5.64159344 | 1.25394008 | 0.31885596 |
| 49 | H | -4.53278930 | 3.34233924 | 2.01512908 |
| 50 | H | -5.21107734 | 2.54331018 | 0.87533100 |
| 51 | H | 6.54778950 | 2.07130014 | -3.20160830 |
| 52 | H | 5.97109846 | 1.67505612 | -1.80346520 |
| 53 | H | -4.85773632 | -1.78195513 | -1.65851819 |
| 54 | H | -4.63112330 | -3.24624624 | -1.09187515 |
| 55 | H | 3.83255531 | 2.03465714 | 1.01095101 |
| 56 | H | 3.54501929 | 0.75843605 | 0.19611695 |
| 57 | H | 0.15657304 | 6.10999243 | 0.41145796 |
| 58 | H | 0.99318510 | 5.91461442 | -0.85789913 |
| 59 | H | 2.37971920 | -2.50920319 | 0.39480696 |
| 60 | H | 3.42684728 | -3.63829527 | 0.94253800 |
| 61 | H | 2.89171924 | -4.82763735 | -0.82535313 |
| 62 | H | 2.92596824 | -6.27622446 | -0.30166809 |
| 63 | H | -3.27179420 | -0.76750006 | 0.53814797 |
| 64 | H | -4.47124829 | -1.68477313 | 0.61658398 |
| 65 | H | -6.24047642 | 0.95431806 | -0.07957307 |
| 66 | H | -4.93940732 | 0.37990202 | 0.59755898 |
| 67 | H | -3.88358525 | -5.29194439 | -2.01124821 |
| 68 | H | -2.89749418 | -4.97976336 | -0.86879913 |
| 69 | H | -4.74280931 | 4.70906833 | -1.90315020 |
| 70 | H | -4.41053129 | 5.20349337 | -3.35614231 |
| 71 | H | 0.99380810 | -0.67123805 | 0.03523893 |
| 72 | H | 0.36460006 | -1.96590315 | 0.56481397 |
| 73 | H | -1.47505108 | -2.48101818 | 1.04799501 |
| 74 | H | -1.18609805 | -3.60078826 | -0.09639107 |
| 75 | H | 1.74536616 | 3.03622321 | 0.51970597 |
| 76 | H | 0.79044009 | 3.38094424 | -0.60895211 |
| 77 | H | 5.57398543 | -3.41426225 | -1.09698815 |
| 78 | H | 4.32992134 | -2.58521519 | -0.77841412 |
| 79 | H | -0.56295301 | 0.30890002 | -1.17126115 |
| 80 | H | 0.42889106 | 1.37118309 | -0.51966210 |
| 81 | H | 0.89508410 | -4.46376933 | -2.82225427 |
| 82 | H | 1.52520214 | -5.69023841 | -2.13578622 |
| 83 | H | -1.35656707 | 6.12305044 | -1.66768919 |
| 84 | H | -2.45061015 | 5.08034036 | -2.09949222 |
| 85 | H | 5.03719339 | -1.36795810 | -2.55973725 |
| 86 | H | 5.16063940 | 0.06640000 | -3.17701630 |
| 87 | H | 4.23931334 | 2.91248120 | -2.64695026 |
| 88 | H | 3.10681625 | 2.13902915 | -2.00236921 |
| 89 | H | -2.58287315 | 2.51037618 | 0.57929697 |
| 90 | H | -2.38902414 | 1.16633908 | -0.19294508 |
| 91 | H | 0.22611305 | 4.44344831 | -2.41394324 |
| 92 | H | 1.79900416 | 4.02583928 | -2.42331324 |
| 93 | H | 2.01554118 | 0.32285502 | -1.60715918 |
| 94 | H | 3.48819628 | -0.03363001 | -2.12646222 |
| 95 | H | -2.91577318 | -0.50202504 | -1.56453118 |
| 96 | H | -1.83963810 | -1.41060811 | -0.73712312 |
| 97 | H | -2.52324715 | 2.24896416 | -1.99784421 |
| 98 | H | -3.43155422 | 2.97439821 | -2.98158128 |
| 99 | H | -3.89014825 | 0.68474804 | -2.81003927 |
| 100 | H | -4.74328531 | 0.42099802 | -1.56575818 |
| 101 | O | 4.82278538 | -2.66806320 | 3.74963620 |
| 102 | O | 2.44265621 | -1.70887513 | 4.35921725 |
| 103 | O | 0.12367404 | -3.21240224 | 3.56813919 |
| 104 | O | -7.90636454 | 2.61747418 | 1.93353907 |
| 105 | O | 0.75328909 | -4.89870236 | 1.45240904 |
| 106 | O | 6.50578450 | -2.50691119 | 1.52996504 |
| 107 | O | 5.64913044 | -5.12048837 | -0.57674511 |
| 108 | O | 4.43638535 | -4.59897434 | 1.91104807 |
| 109 | O | 2.80305023 | -6.77436049 | 1.33297303 |
| 110 | O | -7.71668052 | -1.98697315 | -0.15043508 |
| 111 | O | -2.76538317 | 3.17918622 | 2.31903810 |
| 112 | O | -1.68798009 | -1.49663511 | 2.54902412 |
| 113 | O | 1.76729616 | 0.33487002 | 2.65284812 |
| 114 | O | -0.82542203 | 1.09467207 | 2.23444409 |
| 115 | O | -8.18826356 | 3.43094924 | -0.75811212 |
| 116 | O | 3.13791826 | 6.20285244 | 1.12220001 |
| 117 | O | -5.84605239 | 4.86420434 | -0.76783512 |
| 118 | O | -7.87200154 | 0.71813305 | -0.05484907 |
| 119 | O | 3.37080727 | 3.61452925 | 0.44230596 |
| 120 | O | -1.14070705 | -5.20719438 | -0.50111410 |
| 121 | O | -1.02748304 | 4.73987434 | 0.94809600 |
| 122 | O | 7.69590058 | -0.93060807 | -0.39604710 |
| 123 | O | 6.40027149 | 1.45595110 | -0.26183109 |
| 124 | O | -5.34895035 | 3.31618423 | 1.47423604 |
| 125 | O | 5.73585144 | 1.76653012 | -2.78061627 |
| 126 | O | -5.09320134 | -2.38895518 | -0.93111213 |
| 127 | O | 3.90811231 | 1.05389707 | 1.06211301 |
| 128 | O | 0.77730509 | 6.58376047 | -0.17838808 |
| 129 | O | 2.89008224 | -3.32988725 | 0.16637594 |
| 130 | O | 3.07403525 | -5.74044542 | -1.12940515 |
| 131 | O | -3.98365926 | -1.02605108 | 1.16240502 |
| 132 | O | -5.25249135 | 1.07639107 | -0.02134307 |
| 133 | O | -3.78599324 | -4.75858935 | -1.21612216 |
| 134 | O | -4.04554926 | 4.67920033 | -2.63468426 |
| 135 | O | 1.17414712 | -1.41206811 | 0.67207898 |
| 136 | O | -1.12616005 | -2.63914520 | 0.14485794 |
| 137 | O | 0.81358709 | 2.86201820 | 0.23258495 |
| 138 | O | 5.26233841 | -2.48283818 | -1.06078114 |
| 139 | O | 0.37833706 | 0.50328003 | -0.99037314 |
| 140 | O | 0.67409908 | -5.34660339 | -2.48295225 |
| 141 | O | -1.53199708 | 5.17629637 | -1.74888719 |
| 142 | O | 4.59638636 | -0.73897306 | -3.16521130 |
| 143 | O | 3.35971927 | 3.06375422 | -2.24456523 |
| 144 | O | -2.07098312 | 2.08289814 | -0.13732408 |

| 145 | O |  1.06448511 |  4.43824731 | -1.92279521 |
| 146 | O |  2.96638024 |  0.50332503 | -1.47113017 |
| 147 | O | -2.18038013 | -0.49678604 | -0.89514613 |
| 148 | O | -2.91158118 |  2.15256715 | -2.88775728 |
| 149 | O | -4.33716928 | -0.05752901 | -2.32733124 |
| 150 |   |             |             |             |